\documentclass[preprint2]{aastex}

\newif\ifmakepreprint

\makepreprinttrue

\shorttitle{Compression of Floating-point Images}
\shortauthors{Pence, White, \&  Seaman}

\begin{document}


\title{Optimal Compression of Floating-point Astronomical Images Without
Significant Loss of Information}


\author{W. D. Pence}
\affil{NASA Goddard Space Flight Center,
    Greenbelt, MD 20771}
\email{William.Pence@nasa.gov}

\author{R. L. White}
\affil{Space Telescope Science Institute, Baltimore, MD 21218}

\and

\author{R. Seaman}
\affil{National Optical Astronomy Observatories, Tucson, AZ 85719}



\begin{abstract}
We describe a compression method for floating-point astronomical images that
gives compression ratios of 6 -- 10 while still preserving the scientifically
important information in the image.   The pixel values are first preprocessed
by quantizing them into scaled integer intensity levels, which removes some of
the uncompressible noise in the image.  The integers are then losslessly
compressed using the fast and efficient Rice algorithm and stored in  a
portable FITS format file.  Quantizing an image more coarsely gives greater
image compression, but it also increases the noise and degrades the precision
of the  photometric and astrometric measurements in the quantized image. 
Dithering the pixel values during the quantization process can greatly improve
the precision of measurements in the images.  This is especially important if
the analysis algorithm relies on the mode or the median which would be
similarly quantized if the pixel values  are not dithered. We perform a series
of experiments on both synthetic and real astronomical CCD images to
quantitatively demonstrate that the magnitudes and positions of stars in the
quantized images can be measured with the predicted amount of precision. In
order to encourage wider use of these image compression methods, we have  made
available a pair of general-purpose image compression programs, called fpack
and funpack, which can be used to compress any FITS format image.
\end{abstract}



\section{Introduction}

The exponential growth in the volume of astronomical data being generated from ever
larger format imaging detectors continues to drive up the data handling costs of both
large and small telescope projects. The cost of archiving the images is only one part
of the problem.  A typical observational work flow involves  a long  cascade of
temporary and permanent copies of the data replicated from observatory to data
processing facility to deep storage site to science archive center to virtual
observatory portal to multiple end users.  Each science image provided as input to a
pipeline will produce several output images as a result of processing  operations
such as resampling onto a standard grid, co-adding, mosaicking, and any analysis steps
specific to the science program. The problem is one of throughput -- not just storage
-- of the total system data flow. Network transmission costs also rival or exceed the
cost of storage media and can be breathtakingly large for spacecraft or remote
mountaintops.  Often there is an upper limit to the network bandwidth  at any price.
Making effective use of the available image compression technologies is an important
component in dealing with these data handling costs. 

Lossy data compression techniques, which do not exactly preserve each image pixel
value but do still preserve the required scientific information content of the image,
are already used by many projects.  To cite just a few examples, the  Kepler space
telescope project \citep{caldwell2010}  uses a lossy image quantization method to
reduce the data volume to fit within the limited bandwidth from its heliocentric
orbit, the GONG helioseismology project \citep{harvey1996, goodrich2004} relies on
lossy compression to transmit  images from its telescopes deployed at six locations
worldwide, and the NOAO High-Performance Pipeline \citep{valdes2007} includes a final
quantization step from an internal floating-point representation to the final integer
archive data products to achieve greater image compression.

In a previous contribution \citep[hereafter, Paper I]{pence2009}, we demonstrated that the
maximum possible lossless compression ratio for integer format astronomical
images is determined by the amount of noise in the background pixels (i.e., the
``sky'')  in the image.  The noise in astronomical images often has 2 main
components: Poissonian-distributed photon noise and Gaussian distributed 
``read-out'' noise.  If each pixel is represented by BITPIX  bits (usually
16 or 32 bits) and if, on average, $N_{bits}$ of those bits are filled with
uncompressible, randomly fluctuating noise, then the maximum theoretical
compression ratio for that image is given by BITPIX /  $N_{bits}$.  In
practice, no compression algorithm is 100\% efficient, so the actual maximum
compression ratio, $R$, is given by 
\begin{equation}
 R = {\sf BITPIX}/(N_{bits} + K)
\label{eq:ratio}
\end{equation}
where $K$ is an empirical measure of the efficiency (or overhead) of the algorithm in
units of bits per pixel.  We compared several lossless compression  algorithms and
found that the  Rice algorithm  \citep{rice1993, white1998}, which has a small $K$ value
of  about 1.2 bits per pixel,  provided the best combination of speed and
compression efficiency. 

The relationship between noise and entropy in images is discussed more fully in
the appendix of Paper I, based on the seminal work by \cite{shannon1948}, where we  
showed that the ``equivalent'' number of
noise bits per  pixel in an image can be calculated from the RMS noise ($\sigma$)
of the pixels in background regions of the image such that
\begin{equation}
N_{bits} = {\log}_2 (\sigma \sqrt{12}) = \log_2 (\sigma) + 1.792 
\label{eq:nbits}
\end{equation}
which, when combined with equation 1, gives
\begin{equation}
 R = {\sf BITPIX}/(\log_2 (\sigma) + 1.792 + K)
\label{eq:ratio2}
\end{equation}

In this current article we extend the previous analysis of integer images to
study compression techniques for astronomical images in floating-point format.
In the cases we are mainly concerned with here, these images originally  had
integer pixel values, but were converted into 32-bit IEEE floating-point format
during the calibration processing (e.g., bias subtraction, flat-fielding,
absolute flux calibration, etc.). Equation \ref{eq:ratio2} applies to floating
point images as well as integer images, however typical floating-point
astronomical images contain so much noise that it is impossible to achieve
significant amounts of compression with lossless algorithms (often less than a
factor of 2).   One explanation for this excess noise is that a 32-bit IEEE
floating-point number can express 7 decimal places of precision but this usually
far exceeds the inherent precision of individual image pixel values.  As a
result many of the least significant bits in the mantissa of the floating-point
pixel values are effectively filled with quasi-random noise which is inherently
uncompressible.  While there are  counter-examples of floating-point FITS arrays
that have very little noise and can be losslessly compressed effectively (e.g.,
generated by theoretical simulations), these are not typical of the  types of 
floating-point images commonly found in large astronomical data archives and are
not the subject of this article.

There are many published articles on lossless floating-point data
compression schemes \citep[see ][and reference therein as recent
examples]{lindstrom2006} but it is beyond the scope here to summarize them
in detail. The main point is that ultimately all of these lossless 
compression techniques face the same Shannon entropy limit, and the only
way to achieve greater compression of noisy floating-point images is to use techniques
that discard some of the noise. These methods are technically ``lossy''
because they do not exactly preserve the pixel values, however, if only
noise is discarded, then the compression can still be considered lossless
from a scientific standpoint because all the useful information is
retained.

In the remainder of this article we describe a lossy compression technique 
for floating-point astronomical images that provides an optimal combination
of speed,  compression ratio, and preservation of information content. It is
faster and achieves much higher compression than generic lossless file
compression algorithms like GZIP \citep{gailly}, yet  it  produces no
significant loss of information in the image when used appropriately. In \S
\ref{s:methods} we describe in detail the quantization and compression
techniques that have been implemented in our publicly available {\em fpack}
and {\em funpack} image compression utility programs. Then in \S
\ref{s:tests} we describe the results of several experiments that demonstrate
that  astronomical floating-point images can be compressed  by up to a factor
of 10 without significant loss of astrometric or photometric precision.
Finally, \S \ref{s:discussion}  summarizes  the results and gives
recommendations for achieving the best compression of astronomical images.

\section{Quantization and Compression Methods}
\label{s:methods}

The most common lossy compression technique for floating-point images is to
preprocess the pixel values by quantizing them into a smaller set of discrete
values prior to applying a lossless compression algorithm.  In the simplest
case, the values are rounded into a grid of equally spaced floating-point
levels. This reduces the number of different bit patterns in the image
pixels  (i.e., reduces the entropy in the image) and improves the
efficiency of file compression  programs like GZIP  which
accumulate a dictionary of the most common bit patterns in the file and
represent them using a shorter code in the compressed file.  
\cite{watson2002}  applied an analogous technique to integer images. In order
to accommodate compression algorithms, like Rice, that only operate on
integer arrays, the quantized floating-point values are usually represented
by scaled integers so that the image pixel values are approximated by
\begin{equation}
{\sf FloatValue} = {\sf ScaleFactor} \times {\sf IntegerValue}  + {\sf ZeroPoint} 
\label{eq:scale}
\end{equation}
Note that this integer scaling technique was the only way to represent
floating-point images in the FITS data format \citep{hanisch2001} before
support for the IEEE  floating-point format was officially added in 1990. 

As an aside, there are many articles in the literature   
\citep[e.g.,][]{nieto1999, gowen2003, nicula2005, seaman2009, bernstein2010} 
that advocate using a square root scaling function, in part because the
Poissonian shot noise scales by this same factor.  What is usually not stated
in these studies is that practically the same increase in compression that
is  obtained after applying a square root scaling function can be obtained 
by linearly scaling {\em all} the pixels in the image by the same factor as
is applied to the background pixels during the square root scaling.  Since
the compression ratio is determined mainly by the noise in the background
areas of the image (from equation \ref{eq:ratio2}), the amount of scaling that is applied
to the  relatively infrequent  bright pixels usually makes little difference
to the overall compression ratio of the image.  In experiments on simulated
astronomical images, \cite{bernstein2010} found that using square-root
scaling only produced significantly better compression than linear scaling
when  more than 10\% of the  image pixels are affected by isolated bright
objects or cosmic rays.

\subsection{FITS Tiled Image Compression Convention}

\cite{white1999} developed the technique that forms the basis of the FITS
tiled-image compression format that is used here along with a few new
refinements.  Each row of the image (or in principle, any other rectangular
``tile'' in the image) is compressed separately to provide fast random access
to individual sections of an image without having to uncompress the entire
image.  In the case of floating-point images, the pixel values are
converted to integers with a scale factor that,  by default, is
proportional to the measured amount of noise in that tile.

The noise in each tile of the image is calculated using a robust algorithm
that was originally developed to autonomously measure the signal-to-noise in
spectroscopic data \citep{stoehr2007}.  In particular, we use their third
order  ``Median Absolute Difference'' (MAD) formula to compute the standard
deviation of the  pixel values:
\begin{equation}
\sigma = 0.6052 \times {\sf median} (  -x_{i-2} + 2 x_i - x_{i+2} ) 
\label{eq:noise3}
\end{equation}
where $i$ is the vector index of the pixel within each row of the image,
and $x_i$ is the value of the $i^{th}$ pixel. The median value is computed
over all the pixels in each row of the image. In the limiting case where
the pixel values have a Gaussian distribution, this formula converges to
same value as the Standard Deviation of the pixels.  Note that this formula
is not affected by linear intensity gradients across the image (which are
canceled out by the first and  third terms), and the use of the median
makes the result insensitive to the presence of outlying  large pixel
values.  For example, if one randomly sets 5\% of the pixels in an image
with otherwise Gaussian-distributed noise to very large values to simulate the
effects of cosmic rays, then the MAD noise estimate only increases by
about 20\%.  

Once the noise level has been calculated, the floating-point pixels are
quantized into scaled integers where the quantized levels are spaced at
some user-specified fraction, q, of the noise, so that the spacing is given
by $\sigma / {\sf q}$.  Normalizing the quantization spacing to $\sigma$ is
a convenient way to produce similar quality compressed images regardless of
the intrinsic noise level in the image.  The scaled integers are then
compressed using the default Rice algorithm, or one of the other optional
compression algorithms.  Finally, the compressed stream of bytes is stored
in a FITS binary table structure as defined in the FITS tiled image
compression convention \citep{pence2000, seaman2007}.

Since the noise $\sigma$ in the array of quantized integers is simply equal
to q, the expected image compression ratio, from equation \ref{eq:ratio2},
is given by
\begin{equation}
 R = {\sf BITPIX}/(\log_2 ({\sf q}) + 1.792 + K)
\label{eq:ratio3}
\end{equation}
For example,  quantizing a 32-bit floating-point image using q values of  1
or 4 will produce compression ratios  of about 10 or 6.4, respectively, when
using the  Rice algorithm that has $K \simeq 1.2$.  If the same image is
stored in 64-bit double precision format it will compress by twice this
amount (i.e., the compressed files will have the same size). In most cases,
the compression ratio does not depend very much on the distribution of
objects or structures within the image itself. Note that this formula tends
to break down for q values much less than 1  because many compression
algorithms become less effective on  very low entropy images and  because
the size of the FITS file header,  which remains uncompressed, becomes
relatively more significant. 

One important caveat with the use of this quantizing method is that if the
noise level in the image is significantly overestimated for some reason
(for our purposes it is generally sufficient if it is accurate to within
about a factor of 2), then  the image may be inadvertently quantized more
coarsely than expected for a given q value, thus possibly causing a loss of
information in the compressed image.  Circumstances where the MAD algorithm
could  overestimate the noise include the following:

\begin{itemize}

\item

If a large fraction of the pixels in an image tile are covered by bright
objects and have values (and noise) which is many times greater than in
the fainter pixels, then the MAD noise estimate will be representative of
those bright pixels and not that of the fainter ``background'' pixels. 
This may cause the fainter pixels to be more coarsely quantized than
would be desired.

\item

If a significant amount of the pixel-to-pixel variation in the image is
real and not due to noise, (e.g., if there are large pixel-to-pixel
variations in the detector sensitivity, or if the observed object
contains significant structure on a pixel size scale), then the noise
level may be overestimated. 

\item

If the noise in an image is anti-correlated between adjacent pixels
(e.g., after certain types of image convolutions) then the noise will be
overestimated.  Note, however, that equation \ref{eq:noise3} is a
function only of the values in every other pixel in each row of the
image, so the anti-correlation scale length must extend over at least 2
pixels to affect the MAD noise estimate. 

\end{itemize}

Our {\em fpack} compression program (see \S \ref{s:fpackfunpack})
offers several options for dealing with this issue.  One simple method is
to just compress the image using a larger q value to compensate for the
possible MAD noise overestimate, although this can negatively affect the
overall compression ratio of the image.  Another option in cases where
only a small fraction of the rows in an image might be affected is to
compress the entire image as a single tile, rather than using the default
row-by-row tiling pattern, so that the MAD noise estimate then more
accurately reflects the noise in the background regions of the image as a
whole.  Finally, {\em fpack} users do not need to rely on the MAD noise
estimate at all, and instead can directly specify  the desired spacing
between the quantization levels. This latter option is especially
appropriate for projects that generate large amounts of relatively 
homogeneous images because it ensures that all the images will be
compressed using the same quantization factor.   This method has
the added advantage that it will improve the compression speed because
it is not necessary to compute the MAD noise value in this case.  

\subsection{Benefits of Dithering}

The effects of linear quantization are naturally greater in the fainter areas of
an image than for the brighter pixels where the Poissonian uncertainty of the
photon  counts can be much larger than the quantized spacing.   Measurement of
the local ``sky'' background around the image of a star or galaxy can be
especially vulnerable to  the effects of quantization, in part because it is
usually necessary to  identify and correct for pixels  that are affected by other
objects or by defects in the detector. 
The issue of detecting small amplitude signals in a quantized system is a
well-studied problem in engineering and communications fields where the phenomenon is
known as  ``stochastic resonance''. The somewhat  counter-intuitive solution for
improving the signal-to-noise of measurements of quantized data is to add a moderate
amount of noise into the system.  When applied to images, this technique is commonly
called ``dithering''.  \cite{widrow2008} devote 2 chapters of their book to the
theory and practice of dithering and recommend  using a clever  ``subtractive
dithering'' technique,  first proposed by \cite{roberts1962}. This technique
overcomes the drawback of having to add noise to the image:  a dither is added to the
quantizer input, and the same dither is subtracted again from the quantizer output. 
The dither thus behaves as a catalyst which makes the process work better  but does
not appear in the output image.   We have adopted this technique when quantizing
floating-point images by adding a random dither,  $R_i$, with a value uniformly
distributed between 0 and 1
during the scaling process,
\begin{equation}
I_i = {\sf round}(F_i / {\sf ScaleFactor} + R_i - 0.5)
\label{eq:dither}
\end{equation}
where $F_i$ is the original floating-point value and $I_i$ is the the
quantized integer value. The interesting trick that distinguishes 
subtractive dithering from ordinary dithering methods is that exactly the {\em same}
random dither value is  subtracted when converting back to the quantized
float value:
\begin{equation}
F_i = (I_i + 0.5 - R_i ) \times {\sf ScaleFactor}
\label{eq:undither}
\end{equation}
The net effect of this subtractive dithering operation is to shift the
entire grid of linearly spaced intensity levels up or down by a random
amount on a pixel by pixel basis.  It should be noted that the dithered
values are uniformly distributed between the quantized levels in this
implementation. One possible future enhancement may be to use a triangular
or Gaussian dither \citep{widrow2008} which may more closely replicate the
actual distribution of pixel values in the image.

In order to use this subtractive dithering method it is necessary to define a
specific pseudo random number generator (PRNG) algorithm for use by both the
compressor and the uncompressor so that the same predictable sequence of random
numbers is used in both cases.  We adopted the PRNG algorithm described by
\cite{park1988} which has been shown to produce statistically independent random
numbers  uniformly distributed between 0 and 1.  However, for pragmatic reasons we
do not compute a unique random number for every single image pixel because (a) it
would  add significant computational overhead relative to the  very efficient Rice
compression algorithm, and (b) true randomness is not required for this purpose
since even crude dithering patterns would still help to mitigate the measurement
biases in quantized images.  Our compromise solution is to calculate a
look-up-table of 10000 random numbers using the above mentioned PRNG, and then
repeatedly recycle through this LUT when calculating the amount of dither for each
pixel in the image.  As a precaution against  introducing  regular cyclical noise
patterns into the image, a pseudo random starting offset is computed each time
when recycling through the LUT.   Finally, 1 out of a possible 10000 initial seed
values for the entire dithering process is computed based on the system clock time
(or optionally the checksum of the  first row of pixel values in the image) to
ensure  that the same dithering pattern is not repeated in every image. This initial
seed value is stored in the header of the compressed image for reuse when
uncompressing the image. 

\subsection{Quantization Noise}

While quantization reduces the entropy in the  representation of the pixel
values (thus improving the image compression ratio), any quantization
operation, even with subtractive dithering,  modifies the pixel values and
thus inherently  increases the RMS noise  in the pixel-to-pixel variations
in the image by an amount given by:
\begin{equation}
\sigma_q^2 = \sigma_0^2 + \Delta^2 / 12
\label{eq:quantizednoise}
\end{equation}
where the total variance $\sigma_q^2$ in the quantized image is the
sum of the variance $\sigma_0^2$ in the original image plus the
quantization noise variance, and where $\Delta$ is the intensity spacing
between the quantized levels.  The factor 1/12 comes from the variance of a
uniform random distribution with unit width
\citep[also see the appendix to Paper I]{janesick2001}. 
Substituting $\Delta = \sigma_0 / q$, the fractional increase in the noise 
caused by quantizing the image
can be expressed as
\begin{equation}
\sigma_q / \sigma_0 = \sqrt{1 + 1 / (12 q^2)}
\label{eq:fractionalnoise}
\end{equation}
Figure 1 shows how the fractional noise ratio (from  equation
\ref{eq:fractionalnoise}) and the compression ratio  (from equation
\ref{eq:ratio3}) of an image containing 
Gaussian-distributed noise both depend on q. 
To foreshadow the results of the experiments that will be described in \S
\ref{s:tests}, this figure shows that q values in the range of 4 to 1
provide a good combination of high compression ratio 
with relatively little added noise.

\begin{figure}[t]
\plotone{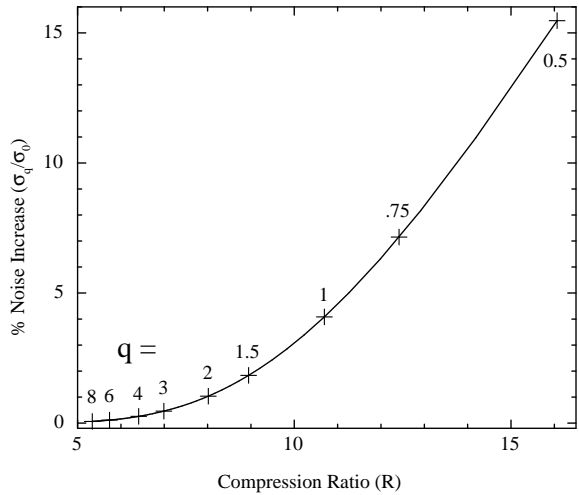}

\caption{Relationship between the compression ratio and the 
fractional increase in the background noise in an image containing
Gaussian-distributed noise for a range of q quantization 
parameter values.
\label{f:fig1}
}
\end{figure}

It should be cautioned that if a floating-point image is repeatedly 
compressed and uncompressed, then the cumulative quantization noise 
will be given by
\begin{equation}
\sigma_q / \sigma_0 = \sqrt{1 + N / (12 q^2)}
\label{eq:multiplenoise}
\end{equation}
where N is the number of compression and uncompression cycles.  
(This assumes that the dithering is re-randomized during each cycle, which is always
the case in our implementation of the subtractive dithering algorithm).
The amplitude of this effect depends strongly on the q value. To put this in
practical terms, an image can be compressed and uncompressed at least 
16 times using q $\geq$ 4 before the noise would increase to the
same level equivalent to compressing the image once with q = 1.  But if the
image is compressed multiple times using q = 1, the noise would increase to 
scientifically unacceptable levels after just a few cycles.  Ideally, an
image should only be compressed once, and then all the  subsequent data
analysis should be performed directly on the tile-compressed FITS file.  If
the software cannot read the compressed format directly, then it should
operate on an uncompressed version of the original compressed file, which
then should not be recompressed.

\subsection{{\em fpack} and {\em funpack} Compression Utility Programs}
\label{s:fpackfunpack}

In order to make the image quantization and compression techniques  that are
described in the previous sections more widely available to the astronomical
community, we have developed a pair of general purpose utility programs, called {\em
fpack} and {\em funpack} \citep{seaman2007}, which can be used to compress and
uncompress any  FITS image in integer or floating-point format.  These utilities
rely on the underlying CFITSIO library \citep{pence1999} to perform the quantization
and compression operations. The {\em fpack} and {\em funpack} utility programs were
used in the experiments that are described in the following sections to quantize and
compress the images.   Further information about {\em fpack} and {\em funpack} is
available from the HEASARC web site at 
{\tt http://heasarc.gsfc.nasa.gov/fitsio/fpack}.

\section{Experiments on Quantized Images} 
\label{s:tests}

In this section we present the results of experiments designed to show how the
measurements of objects in an image are affected as the image is quantized by varying
degrees.  In particular, we will verify that the noise in the image increases as a
function of q by the amount predicted by equation \ref{eq:fractionalnoise}, and more
significantly, that the statistical errors on the magnitude and position 
measurements of faint objects in an image, which are limited mainly by the background
noise, also increase by a similar factor. These results will provide general
guidelines for achieving the greatest amount of image compression while still
preserving the required level of scientific precision in the image.

Section \ref{s:setup} describes the method of constructing  the
simulated CCD images that are used in the first 2 experiments.  The first experiment, in
\S \ref{s:exp1}, examines how quantization affects the uncertainties
of measurements of single star images, and the second experiment, in \S
\ref{s:exp2}, examines the case where many quantized images are added together
to detect sources far below the detection threshold of a single image.
Finally, the third experiment, in \S \ref{s:exp3}, is performed  on a set of
actual astronomical images to verify the results obtained from the synthetic 
images.

\subsection{Data Model and Measurement Methods}
\label{s:setup}

In order to determine how quantization affects the precision of measurements of
objects in an image, we generated a large sample of realistic CCD star images with
known input positions and magnitudes.   This allows us to precisely  calculate the
errors on the measured positions and magnitudes in the quantized images.  All the
stars have circular Gaussian profiles with $\sigma = 1.0$ and FWHM = 2.35 pixels. 
This is typical of the spatial resolution commonly found in astronomical CCD images
and is adequate to avoid the difficulties when analyzing spatially undersampled
images. The central location of the star images, relative to the pixel grid, was
varied so as to average out any subtle biases in the subsequent star
detection and  measurement steps that might depend on the exact position.  The total
integrated flux in the stars covered a range of 10 magnitudes (a factor of 10000 in
intensity) in 0.5 magnitude increments.  Finally, the sky background was simulated by
adding 1000 counts to each pixel.

Poissonian-distributed shot noise and Gaussian distributed ``read-out'' noise was
then added to each of these star images to simulate real CCD images. The shot noise
in each pixel was randomly calculated using a $\sigma$ equal to the square root of
that pixel value (which implicitly  assumes that the ``gain'' of the simulated CCD
has been set to 1 electron per analog-to-digital readout count), and the readout
noise was calculated using $\sigma$ = 10.  The read-out noise in  these images is
relatively small compared to the shot noise in the sky background, which is usually
the case for real astronomical CCD images that have a moderately bright background
level. The total noise in the background areas of these images has $\sigma =
\sqrt{1000 + 10^2} = 33.2$.   Different starting random seed values were used so that
the actual noise distribution varies in every image.  

The widely used SExtractor source extraction program \citep{bertin1996} was employed
to objectively detect and measure the position and magnitude of the stars in these
simulated, noisy CCD images. SExtractor calculated the positions  (given by the
X\_IMAGE and Y\_IMAGE output parameters) from the flux-weighted centroid of all the
pixels in each star image above an empirically determined flux detection threshold. 
The total magnitude of each star  (given by the MAG\_APER parameter) was measured
within a 7-pixel diameter aperture (i.e., out to 3.5 $\sigma$ of the Gaussian
profile)  around the centroid position.   We calculated the actual errors on these
measurements  from the differences between the known input position and the input
flux (corrected for the 0.3\% of the flux that falls outside the 3.5$\sigma$
aperture) of each star in the synthetic image. For reference, the faintest
stars that SExtractor could reliably detect in these images contained about 1000 net
counts, which is coincidentally equal to the mean sky counts in  each pixel.  We
arbitrarily set the zero point of the instrumental magnitude scale,  
\begin{equation}
    m = -2.5 \log ({\sf flux}) + {\sf ZeroPoint}
\end{equation}
so that these faintest detected stars have a magnitude of 20.0. 

We repeated the SExtractor measurements of the simulated CCD images after
quantizing  and compressing each image with {\em fpack} using q values ranging from 8
to  0.25, both with and without the subtractive dithering option. Since the
SExtractor program cannot directly read images in the compressed FITS format, it was
necessary to convert them back into the standard FITS image format using {\em
funpack} (but the pixel values remain quantized, of course). In the case where
dithering was not applied, we also selected the option to compress the entire image
as a single large tile so that all the pixels are quantized into the same fixed grid
of intensity values. If we had used the default row-by-row tiling option instead, it
effectively  would have introduced some dithering of the quantized levels between
adjacent rows, and the results would be intermediate between the dithered and
non-dithered cases presented below.

\begin{figure}
\plotone{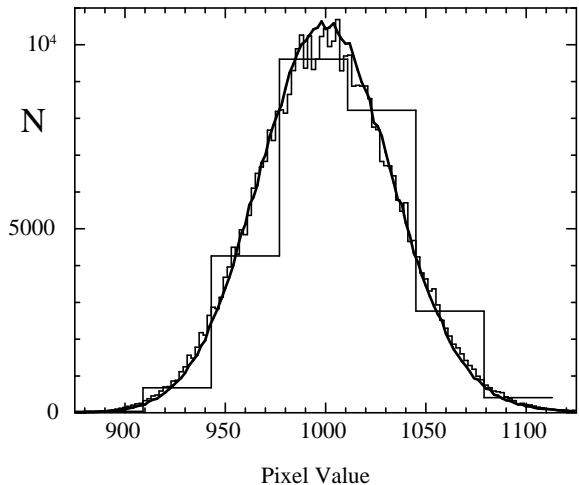}

\caption{Histograms of the pixel intensity distribution in one of our test
images which
demonstrates the beneficial effects of subtractive dithering when quantizing
the image with q = 1. When dithering is applied (the finely stepped
histogram), the pixel distribution closely matches that in the original image
(the continuous curve).  When dithering is not applied (the coarsely
stepped histogram) then subtle gradations in image intensity are lost.
\label{f:fig2}
}
\end{figure}

\begin{figure}
\plotone{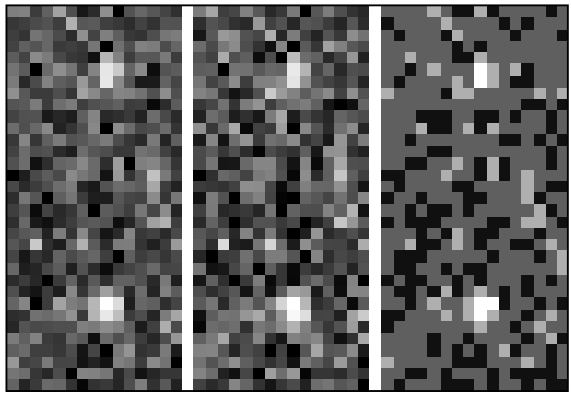}

\caption{Images of a pair of faint (m = 20) artificial stars (left panel)
and the same image after quantizing the image with q = 0.5 with dithering
(middle panel) and without dithering (right panel).  Without dithering, the 
background breaks up into broad patches of constant intensity
level, and the star images become harder to detect.
\label{f:fig3}
}
\end{figure}

Figure 2 shows the dramatic effect that subtractive dithering has on the  histogram
of the pixel values in a q = 1 quantized image; the dithered image histogram is
nearly identical to the that of the original image, whereas the non-dithered
histogram clearly shows the coarse intensity binning.  Without dithering, the
image breaks up into  discrete ``bands'' of constant intensity which makes it more
difficult to detect faint features in the image. This effect is shown in  Figure 3,
(using an even more coarsely quantized q = 0.5 image to enhance the effect), where
the majority of the background pixels all have the same (medium gray) intensity
value. 

\subsection{Experiment 1: Single Images}
\label{s:exp1}

In this first experiment, we examine how the uncertainties of the photometric and
astrometric measurements of individual stars  depend on the q quantization
factor.  To measure this effect, we computed the Standard Deviation of the 
magnitude and position errors (i.e., the value computed by SExtractor minus the
known input magnitude or position value) for 6250 simulated star images at each 0.5
magnitude increment.  Figure 4 shows how the magnitude  uncertainties decrease as
the brightness of the star increases (towards the right).  The lower, thicker line
in the figure was derived from the original, unquantized images and shows that the
statistical uncertainty on the magnitudes decreases from $\sigma = 0.23$ for the
faintest detectable stars (with m = 20), to  $\sigma = 0.018$ for stars that
are 3 magnitudes brighter. The line has a slope close to 1.0 (in log -  log
coordinates) as expected in the limiting case where the noise is dominated by the
sky, and hence the signal-to-noise ratio of the measurement increases in direct
proportion to the signal.  

The other 3 lines in Figure 4 were derived from the same images after they were
first quantized and compressed with successively coarser q values of 1.0, 0.5, and
0.25, when also applying the subtractive dithering option. The vertical displacement of
these lines shows that  the measurement errors are systematically larger in the
quantized images as compared to the measurements in the original image.  This
increase is relatively small for q $\geq$ 1, but increases rapidly for coarser
quantization values.  

Figure 5 shows a similar plot of the Standard Deviation of the stellar centroid
measurements, in units of pixels, as a function of the magnitude of the star and
the q  quantization factor.  In the original unquantized images,
the positional uncertainty decreases from $\sigma = 0.24$ pixels for the m = 20
stars, to $\sigma = 0.035$ pixels for the stars that are 3 magnitudes brighter.
The positional measurement uncertainties are larger in the quantized image by
about the same factor as the increase in the magnitude uncertainties shown
in Figure 4.

\begin{figure}
\plotone{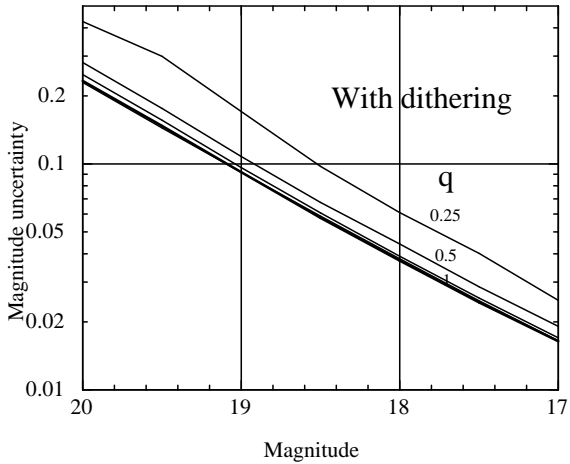}

\caption{The effect of q on the measured magnitude uncertainties when dithering
is applied.
\label{f:fig4}
}
\end{figure}

\begin{figure}
\plotone{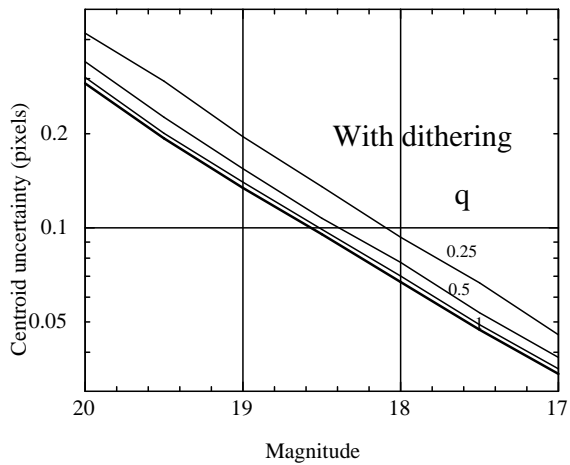}

\caption{The effect of q on the measured positional uncertainties when 
dithering is applied.
\label{f:fig5}
}
\end{figure}

In order to better quantify the effects shown in these 2 figures, Table 1 
summarizes how the noise and the measurement uncertainties  increase as the images
are quantized more coarsely.  The measured percentage increase in the MAD noise
level  in the quantized images is given in column 2 and agrees exactly with the 
predicted value from equation \ref{eq:fractionalnoise}.  More strikingly, the 
magnitude and position uncertainties for the faintest stars also increase by about
the same factor as the noise, as shown in columns 3 and 4.   This means that
equation \ref{eq:fractionalnoise} also provides a good estimate of how much the
measurement uncertainties of the faintest objects in an image, which are limited
by the background noise, will  increase when the image is quantized.  

Finally, columns 5 and 6 in Table 1 give the corresponding increase in measurement
uncertainties for stars that are  5 magnitudes, or a factor of 100, brighter than
the image detection threshold.   As expected, these brighter objects are
relatively less affected by quantization because the inherent  Poissonian noise in
the brighter pixels is larger than the spacing between the quantized levels. 
Also, the small formal statistical errors on the magnitudes and positions of the
brighter stars  (which are less than 0.001 of a magnitude or pixel, respectively,
in this case) are often insignificant  compared to the systematic errors in the
absolute calibration of the measurements.

\begin{deluxetable}{cccccc}
\tablecaption{Increased noise and measurement uncertainties as function of q
\label{t:qnoise}
}
\tablehead{
\colhead{q} & \colhead{$\sigma_q/\sigma_0$} & \colhead{$\Delta{\sf mag}_2$} & \colhead{$\Delta{\sf pos}_2$} & \colhead{$\Delta{\sf mag}_5$}  
& \colhead{$\Delta{\sf pos}_5$ }}
\tablewidth{0pt}
\startdata
4    &  0.26\%   &   0.31\%  & 0.18\% & 0.10\%  & 0.20\% \\
2    &  1.04\%   &   1.1\%   & 0.93\% & 0.25\%  & 0.83\% \\
1    &  4.08\%   &   5.6\%   & 4.1\%  & 1.7\%   & 3.4\% \\
0.5  &  15.5\%   &   19\%    & 15\%   & 5.8\%   & 12\% \\
\enddata
\end{deluxetable}

\begin{figure}
\plotone{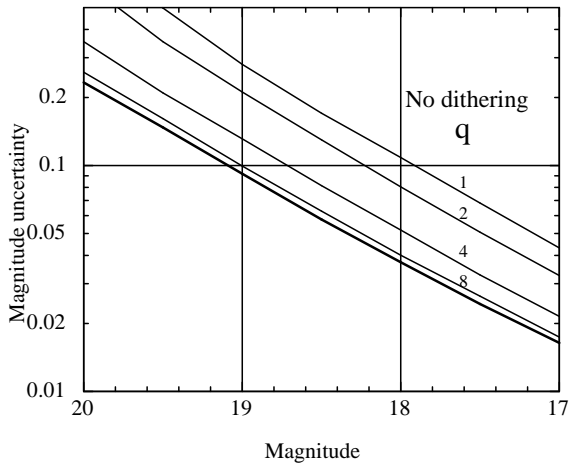}

\caption{Same as Figure 4 but without dithering.  The 
magnitude uncertainties are much larger than in the dithered case.
\label{f:fig6}
}
\end{figure}

\begin{figure}
\plotone{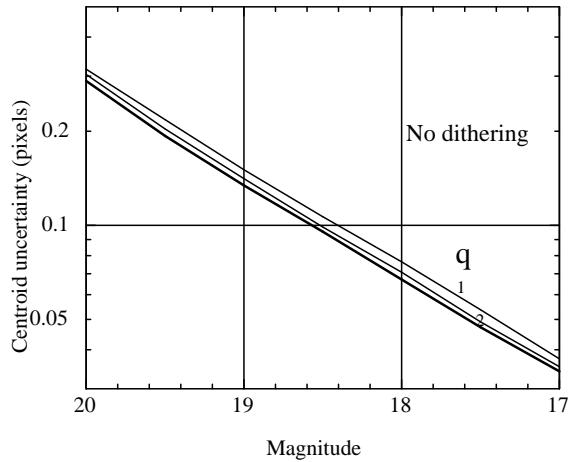}

\caption{Same as Figure 5 but without dithering. 
\label{f:fig7}
}
\end{figure}

For comparison, Figures 6 and 7 show that if the quantized pixels are not
dithered then the measurement errors are larger than in the  dithered case
shown previously in Figures 4 and 5.  It is striking that while dithering
provides a modest increase in the accuracy of the centroid measurements, it
greatly improves the magnitude measurements. Further investigation has shown
that the larger magnitude errors are  almost entirely caused by a dramatic
increase  in the errors in the SExtractor background estimate around each
star if the quantized image is not dithered. SExtractor  assumes that the
best estimate of the true background level is given by the peak in the
histogram of the pixel values (i.e., the mode) and it uses the following 
empirically derived formula
\begin{equation}
{\sf mode} = 2.5 \times {\sf median} - 1.5 \times {\sf mean}
\label{eq:mode}
\end{equation}
to correct for the slight skewness in the pixel distribution (i.e., an
excess of brighter pixels) that is commonly  seen due to contamination by
other objects in the image. This formula generally works well if the pixel
values have a continuous distribution, as demonstrated in our tests on
the dithered star images, however, it behaves poorly if the pixels are
quantized  because then the median value is also quantized. The amplitude
of the error in the background estimate depends on how  closely one of
the quantized levels happens to match the true background level.  In the
worst case, where the true value lies half way between 2 of the quantized
levels (and assuming that the mean value lies very  close to the true
background level, which is usually the case in our images),  then the
error is equal to  1.25 times the intensity spacing between adjacent
quantized levels. On average, the RMS error in the background estimate
introduced by this effect will be $ (2 \times 1.25) / \sqrt{12} = 0.72 $
times the quantized spacing.  It is ironic that equation \ref{eq:mode} is
intended to provide a more accurate background  estimate than simply
using the median value alone, but it actually increases the errors by a
factor of 2.5 if the image is significantly quantized.

It should be emphasized that while this particular problem applies
specifically to the way SExtractor estimates the background intensity, 
it is likely that similar problems would arise with other astronomical
data analysis packages unless they are specifically designed to deal with
quantized data.  Another way to view this problem is that many algorithms
are not well behaved if the noise is under-sampled. Dithering helps to
eliminate this problem because it preserves the natural distribution of
the pixel values, as shown in  Figure 2.

As a final test,  it is important to establish that there are no subtle
systematic zero-point biases in the measured magnitudes  in the quantized
images.  This is demonstrated in Figure 8, which plots the mean magnitude
residual (true magnitude minus measured magnitude) for the 6250 simulated
star images within each 0.5 magnitude bin after quantizing the images with
a range of different q values.  The upper panel shows the mean residuals
in the subtractively  dithered images, and the lower panel shows the
residuals in the non-dithered quantized images. The thicker line in both
panels was derived from the original unquantized images and shows that
there is  a small intrisic negative bias (-0.02 mag) in the SExtractor
magnitudes  of the faintest stars.  The cause of this small measurement
bias, which is less than 1/10th the statistical error on a single
magnitude measurement, is unknown, but it might be due to the algorithms
that SExtractor uses to measure the fainter stars, or perhaps due to the
particular set of SExtractor configuration parameters that we used in these
tests. The important point here is that a very similar pattern of residuals
is seen (in the upper panel) in the coarsely quantized and dithered images,
which shows that there are no  significant additional magnitude biases in
these quantized images. For $q \ge 2$, the mean residuals match those in
the unquantized images so closely that the points lie within the thickness
of the line in the figure. If dithering is not applied, however, then
significant biases in  the magnitude residuals may be present in the
quantized images, as shown in the lower panel of Figure 8. This effect is
entirely due to the systematic errors caused by the use of  the median
function in estimating the background intensity level, as previously
discussed.

\begin{figure}
\plotone{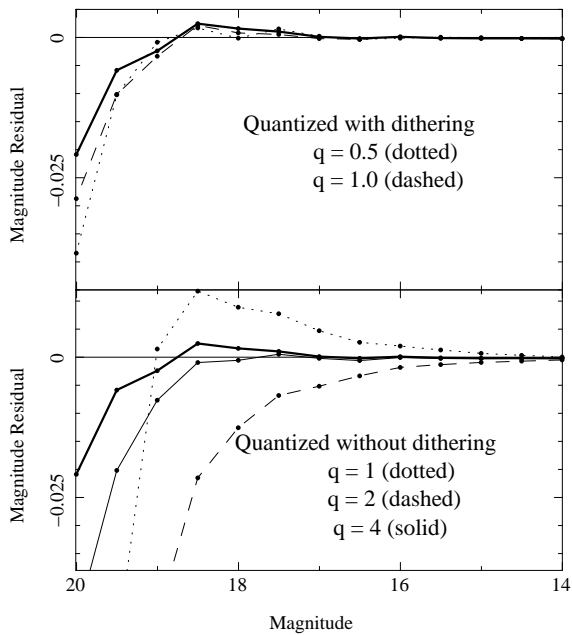}

\caption{
The mean magnitude residual (true magnitude minus measured magnitude)
derived from the simulated star images after quantizing the images with
different q values.   The upper panel shows the case where the pixel values
were subtractively dithered and the lower panel shows the case with no
dithering. The bold line in both cases shows the small inherent bias in the SExtractor
magnitudes in the original images, without any quantization. The lines for q = 2
and q = 4 are hidden under this bold line in the upper panel.
\label{f:fig8} } \end{figure}

\subsection{Experiment 2: Co-added Images}
\label{s:exp2}

The second experiment investigates how well faint simulated Gaussian
stars (as described in \S \ref{s:setup}) that are below the detection
threshold of a single image can be measured after co-adding 250 similar
images together.   To first order, one would expect that this experiment
should detect stars that are about $2.5 \log{\sqrt{250}} = 3.0$
magnitudes fainter than in the single images. For comparison, we also
generated quantized co-added images by quantizing  each of the 250 
individual images (using a range of q values, with and without
subtractive dithering) before  co-adding them.    The magnitude and
position of the stars in each of these co-added images  were then
measured with the SExtractor program,  just as in the first
experiment.   
\begin{figure}
\plotone{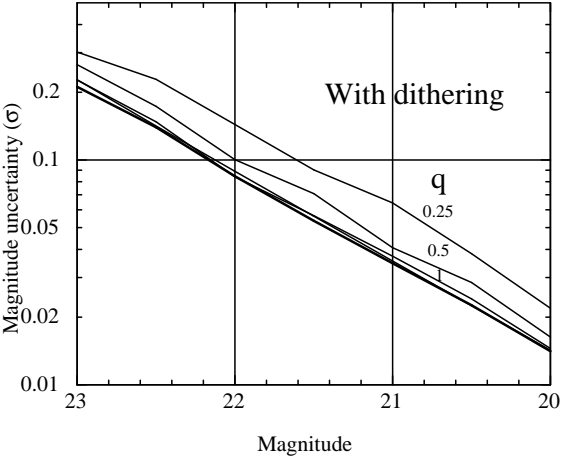}

\caption{The effect of q on the measured magnitude uncertainties 
in the sum of 250 images, with dithering.
\label{f:fig9}
}
\end{figure}

\begin{figure}[htb]
\plotone{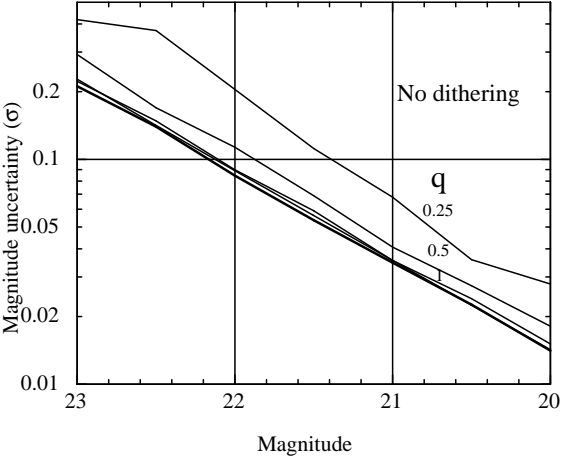}

\caption{Same as Figure 9, but without dithering.  Dithering
the 250 individual images has little effect on the precision of
the magnitude measurements in the summed image.
\label{f:fig10}
}
\end{figure}

Figure 9 shows the results of the photometric measurements in the co-added
images. Other than the fact that the lines in the plots
exhibit more statistical scatter,
since they were derived from a sample of only 250 star measurements in each 0.5
magnitude bin instead of 6250, these results are very similar to the results
from the single images shown in Figure  4,  after allowing for the expected
shift of the horizontal axis by 3  magnitudes. This similarity confirms that
the photometric measurements in the co-added images have the same dependence on
q as in the individual images.  In particular, it confirms that information
about faint stars that are well below the detection threshold in a single
image is still preserved in the quantized images and can be detected
after co-adding many quantized images together.

We also confirmed that the astrometric precision in the  co-added images shows
the same dependence on q as the measurements in the single images. A plot of
the positional uncertainty as a function of magnitude (not shown here)  is
almost identical to Figure 5, except for the 3 magnitude shift of the
horizontal axis. 

Finally, Figure 10 shows the magnitude uncertainties when the quantized images are
not subtractively dithered before co-adding them.   Unlike the case for single
images (as compared in Figures 4 and 6),  dithering has little effect on  the
precision of the photometry in the co-added image.  This is because the co-adding
process effectively introduces its own form of pixel dithering if  the quantized
levels in the different images are randomly offset with respect to each other. 
Thus, there is little added benefit by also dithering the pixel values within each
image.  \cite{price2010} demonstrate an even more extreme case where accurate
photometry and astrometry could be performed on a co-added sum of 1024 images,
each quantized with q = 1/16 = 0.0625 without any dithering.

\ifmakepreprint
\clearpage
\fi

\begin{figure}[ht]
\plotone{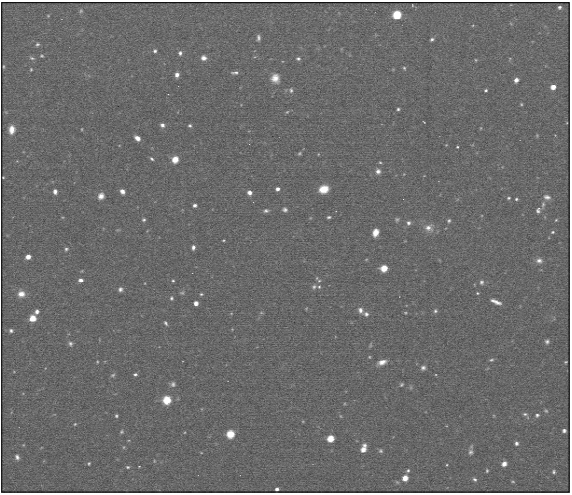}

\caption{Sample region in one of the test images taken with the 
Isaac Newton Telescope at La Palma.  This shows about 10\% of the
total image area.
\label{f:fig11}
}
\end{figure}

\subsection{Experiment 3: Real Astronomical Images}
\label{s:exp3}

In the third and final experiment we performed tests on a set of real astronomical CCD
images to confirm the previous results derived from synthetic images.  We used a sequence of
20 similar CCD images taken with the 2.5-m Isaac Newton Telescope at La Palma  that show a
random distribution of faint stars and galaxies (Figure 11).   All the exposures were for
600 seconds through a V-band filter with the same pointing on the sky  ($12^h 51^m,
26^{\circ} 24'$). These images are publicly available through the virtual observatory portal
at {\tt http://portal-nvo.noao.edu}.

We performed 2 tests to measure the effects of quantization on these images.  In the first
test we coarsely quantized one of the images using q = 1 (with subtractive dithering),  by
compressing it with {\em fpack} and then uncompressing it again with {\em funpack}.  As
predicted by  equation \ref{eq:ratio3}, {\em fpack} achieved a compression ratio of about
10, which is equivalent to 3.2 bits per pixel in the compressed image.  Also as expected
from equation \ref{eq:fractionalnoise}, the measured background noise level increased by
4.1\%, from $\sigma$ = 22.79 in the original image to $\sigma$ = 23.73 in the quantized
image.

We then compared the SExtractor magnitude measurements in the original image to those
derived from the quantized  image.   Since we do not know the true magnitudes of the stars
in this image (unlike in the experiments on the synthetic stars), we can only  compare the
measurements of the stars in the 2 images, both of which have measurement uncertainties.   
The top panel of Figure 12 shows the difference between the 2 magnitude measurements for
each star as a function of the magnitude of the star, and the middle panel shows the
corresponding  {\em relative} errors (the magnitude difference divided by the statistical
error on that magnitude as calculated by SExtractor).  The larger points in these panels
show the mean difference averaged over 0.5 mag bins, which demonstrate that there is no
significant  systematic bias between the 2 sets of magnitudes measurements. The RMS value of
all the relative errors in the middle panel is $0.34 \sigma$.  This is slightly larger than 
the factor of $\sqrt{1/12} = 0.29$ that one would expect simply from the added quantization
noise and may be due to other residual  sources of noise in the SExtractor calculations.  
The fact that this is much less than the approximately $ 1 \sigma$  dispersion that one
would expect when comparing the magnitudes derived from 2 identical CCD images of the same
stars, confirms that quantizing the image with q = 1 has not  introduced  statistically
significant differences in the magnitude measurements. Finally, the lower panel plots the
ratio of the magnitude errors in the quantized and original images which shows that the
statistical errors of the faintest stars are on average about 4\% greater in the q = 1
quantized image, as expected.

\begin{figure}
\plotone{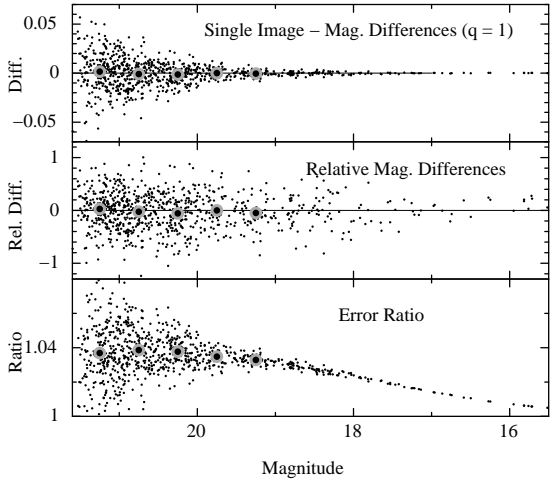}

\caption{Differences in star magnitudes measured in a CCD image
and a q = 1 quantized version of the same image.  The upper panel
shows the absolute magnitude differences, and the middle panel shows
the relative differences after dividing by the statistical error.
The lower panel shows the ratio of the statistical errors in the
quantized image relative to those in the original image.
\label{f:fig12}
}
\end{figure}

\begin{figure}
\plotone{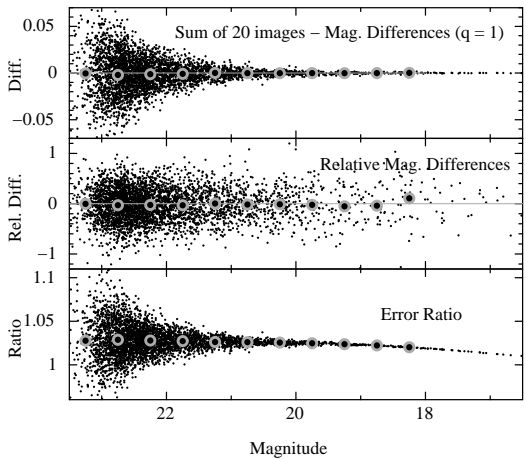}

\caption{Same as Figure 12, except the magnitudes are measured in the
sum of 20 CCD images and in the sum of the q = 1 quantized version of each 
of the images.
\label{f:fig13}
}
\end{figure}

For the second test we  created 2 new images by co-adding the 20 original CCD images  and by
coadding the q = 1 quantized version of each of the images and then repeated the same tests
as described above.  The results, shown in Figure 13, are similar to those in Figure 12
except that the faintest detected stars  are now  $2.5 \log{\sqrt{20}} = 1.6$ mag fainter,
as a result of co-adding the 20 images. The RMS value of the relative errors shown in the
middle panel is $0.33 \sigma$ in this case, which again demonstrates that quantizing the
images with q = 1 has not produced any significant photometric errors, even in objects that
are fainter than the detection threshold in a single image.


\section{Summary and Discussion}
\label{s:discussion}

The main purpose of this work has been to find a more effective compression method for 
floating-point astronomical images than is provided by the lossless methods that are
commonly used (such as GZIP). These floating-point images typically do not compress well
with lossless algorithms  because a large fraction of the bits in each pixel
value representation contain no significant information and are effectively filled with
uncompressible noise. We have adopted the  method of eliminating some of this noise by
quantizing the pixel values into a set of discrete, linearly spaced intensity levels, which
are represented by scaled integer values.  The scaled integers can then be efficiently
compressed using the very fast Rice algorithm.   In order to make these compression
techniques more widely available, we have produced a pair of  utility programs called {\em
fpack} and {\em funpack} that can be used to compress any FITS format image.

For convenience, we define a quantization parameter, q, which is equal to the measured RMS
noise in background regions of the image divided by the spacing between the quantized
intensity levels. Given a particular q value when quantizing and compressing an image, one
can calculate from fundamental principles the expected image compression  ratio, as given by
equation \ref{eq:ratio3}.  Coarser quantization (i.e., smaller q values) gives greater image
compression, but at the same time it increases the RMS pixel-to-pixel noise by an amount
given by equation  \ref{eq:fractionalnoise} which also tends to degrade the precision of the
magnitude and position measurements of the objects in the image.

Our series of experiments on simulated and on real astronomical CCD images
demonstrate that the noise equation \ref{eq:fractionalnoise} also gives a good
estimate of the increase in measurement uncertainties for objects near the
detection threshold in a quantized image (which are noise limited).   For many
practical applications, q values between 1 and 4 provide a good combination of
high compression and low increased noise: q = 1 gives a compression ratio of
about 10 while increasing the noise and the measurement uncertainties  by
about 4\%, and q = 4  gives a compression ratio of 6 with a negligible 0.26\%
increase in the noise and measurement uncertainties. In ``quick-look'' types of
applications, where high scientific accuracy is not of primary importance,
even greater compression can be obtained  by using  q values less than 1.

One necessary requirement for achieving these results, at least when using the
SExtractor program to perform the measurements, is that  the pixel values must
be dithered to be able to accurately measure the background intensity level
around each object.  The algorithm that SExtractor uses to estimate the
background relies on the median of the pixels values, which is not a reliable
statistic if the values are quantized.  Since other astronomical image analysis
software may have similar issues dealing with quantized images, it would be
prudent to always dither the pixel values during the quantization  process to
avoid possible complications later on, especially since dithering introduces no
significant undesirable side effects.

Several other recent studies have reached similar conclusions about the
benefits  quantizing astronomical images to reduce the amount of noise and
improve the  compression ratio. \cite{price2010} performed photometric and
astrometric measurements on stars in simulated CCD images (using a slightly
different methodology than used here) and concluded that no significant errors
were introduced if the image is quantized (without dithering) with q $\geq 2$.
\cite{bernstein2010} found that using q = 1, along with square-root scaling 
does not induce any significant bias in weak-lensing shape measurements of
galaxies  (i.e., the second statistical moment of the image) in simulated
images from future space-based imaging missions. \cite{caldwell2010}  describe
how the Kepler mission is performing on-board quantization of the images with
q values as low as 1.15 to obtain the necessary amount of compression of the
downlinked telemetry while still preserving the required high precision in the
differential photometry measurements of stars. While we recognize that
applying any sort of lossy compression scheme to scientific data  tends to go
against the inclination of scientists and data archive professionals,
hopefully the results of the experiments performed here and in the other cited
references will help to alleviate these concerns.  

While we believe our test results should be applicable to a wide range of
floating-point astronomical images, we strongly encourage users to perform
their own tests to verify that this compression technique is appropriate for
their own data.  Users can easier create a quantized and dithered version of
any floating-point FITS image by compressing it with {\em fpack} and then
uncompressing it with {\em funpack}.  This image can then then be processed
just like the original image to see if there are any significant differences
in the results.



\acknowledgments

We wish to thank the (second) anonymous referee for a thorough review that  
improved the paper and helped us identify the reason for large  
SExtractor magnitude errors when dithering is not used.

\end{document}